\documentclass[twocolumn]{jetpl}

\usepackage{amssymb}
\usepackage{amsmath}
\usepackage{amsfonts}
\usepackage{graphicx}
\usepackage{bm}
\usepackage{color}
\usepackage{multirow}
\usepackage{cite}

\DeclareMathOperator{\Tr}{Tr}

\begin{document}

\title{Helical edge transport in the presence of a magnetic impurity}

\rtitle{Helical edge transport in the presence of a magnetic impurity}

\sodtitle{Helical edge transport in the presence of a magnetic impurity}

\author{P.\,D.\,Kurilovich$^{1,2,3}$, V.\,D.\,Kurilovich$^{1,2,3}$, I.\,S.\,Burmistrov$^{3,1,4,5}$\thanks{e-mail: burmi@itp.ac.ru.}, M.\,Goldstein$^{6}$}

\dates{\today}{*}

\address{
$^{1}$ Moscow Institute of Physics and Technology, 141700 Moscow, Russia \\
$^{2}$ Skolkovo Institute of Science and Technology, 143026 Moscow, Russia \\
$^{3}$ L. D. Landau Institute for Theoretical Physics RAS, 119334 Moscow, Russia \\
$^{4}$ Institut f\"ur Theorie der kondensierten Materie, Karlsruhe Institute of Technology, 76128 Karlsruhe, Germany \\
$^{5}$ Institut f\"ur Nanotechnologie, Karlsruhe Institute of Technology, 76021 Karlsruhe, Germany \\
$^{6}$ Raymond and Beverly Sackler School of Physics and Astronomy, Tel Aviv University, Tel Aviv 6997801, Israel
}

\abstract{We consider the effects of electron scattering off a quantum magnetic impurity on the current-voltage characteristics of the helical edge of a two-dimensional topological insulator. We compute the backscattering contribution to the current along the edge for a general form of the exchange interaction matrix and arbitrary value of the magnetic impurity spin.  We find that the differential conductance may exhibit a non-monotonous dependence on \color{black} the voltage with several extrema. 
}
\maketitle

\textsf{Introduction.}\ ---  Two-dimensional topological insulators (2D TIs) are in the focus of recent interest due to existence of two helical edge states inside the band gap~\cite{Qi-Zhang,Hasan-Kane}. Because of spin-momentum locking caused by strong spin-orbit coupling, electrical current transfers helicity along the edge \cite{Kane-Mele, BHZ}. This ``spin'' current is a hallmark of 
the quantum spin Hall effect{, and it} has been detected experimentally in HgTe/CdTe quantum wells \cite{Konig2007,Roth2009,Gusev2011,Brune2012,Kononov2015}. 
If only elastic scattering is allowed,~and in the absence of time-reversal symmetry breaking, the helical state is a realization of the ideal transport channel with conductance of $G_0=e^2/h$.
This prediction was questioned by the experiments 
in HgTe/CdTe \cite{Konig2007,Nowack,Grabecki,Gusev2014} and 
InAs/GaSb \cite{RRDu1,RRDu2} quantum wells.
Therefore, studies of mechanisms which can lead to the destruction of the ideal helical transport are important. 

A local perturbation breaking the time-reversal symmetry, e.g., a classical magnetic impurity, leads to back\-scatte\-ring of helical edge states and reduction of the edge conductance~\cite{Maciejko2009,Tanaka2011}.
Electron-electron interactions along the edge can promote edge reconstruction and, consequently, spontaneous time-reversal symmetry breaking at the edge \cite{Meir2017}.
Furthermore, even in the absence of time-reversal symmetry breaking, electron-electron interactions 
may induce backscattering \cite{Xu2006}, resulting in the suppression of the helical edge conductance at finite temperatures (see \cite{Gornyi2014} and references therein). 
A combination of electron-electron interactions and magnetic impurities can significantly modify the picture of ideal helical edge transport \cite{Maciejko2012,Yudson2013,Yudson2015,Glazman2016,Yudson2017}. 

{In the absence} of electron-electron interactions along the edge, the ideal transport along the helical edge may still be affected (at finite temperatures) by its time-reversal symmetric interaction with a ``quantum impurity'', that is, an impurity which has its own quantum dynamics, e.g. a charge puddle that acts as an effective spin-1/2 impurity \cite{Goldstein2013,Goldstein2014}, or a  quantum magnetic impurity with spin $S=1/2$ \cite{Maciejko2009,Tanaka2011} or $S\geqslant 1/2$ \cite{Cheianov2013,Kimme2016}. 


In this Letter we study theoretically a modification of the ideal current-voltage characteristics of the {helical edge}  in 2D TI by \emph{weak} scattering off a single magnetic impurity. As a physical realization of such system we have in mind 
the (001) CdTe/HgTe/CdTe quantum well (QW) with a Mn impurity that possesses spin $S=5/2$. Contrary to the previous works, we consider a general structure of the matrix describing exchange interaction between edge states and magnetic impurity [cf. Eq. \eqref{eq: JijH}]. For $S=1/2$ we find an analytical expression for the backscattering current at arbitrary voltage. For larger spin, $S>1/2$, we obtained analytical expressions for the backscattering current at low and high voltages. 

\textsf{Model.}\ --- The low-energy physics of electron and hole states in 2D TI based on the (001) CdTe/HgTe/CdTe quantum well (QW) is described by Bernevig-Hughes-Zhang (BHZ) Hamiltonian \cite{BHZ}. This Hamiltonian is a $4\times 4$ matrix in the basis of electron and heavy-hole states $\vert E_1,+\rangle$, $\vert H_1,+\rangle$, $\vert E_1,-\rangle$, $\vert H_1,-\rangle$, with elements which are linear and quadratic functions of the momentum $\bm{k} = \{k_x,k_y\}$. For the sake of simplicity, we neglect quadratic terms and consider 
simplified version,
\begin{equation}\label{Eq: BHZ-lin}
H =
\begin{pmatrix}
h(\bm{k}) & 0 \\
0 & h^T(-\bm{k}) 
\end{pmatrix} ,
\,
h(\bm{k})= \begin{pmatrix}
M&Ak_{+} \\
Ak_{-}&-M
\end{pmatrix} ,
\end{equation} 
where $A$ and $M$ are material parameters which depend on the thickness of the QW, $k_\pm=k_x\pm i k_y$, and the superscript $T$ denotes matrix transposition. The Hamiltonian \eqref{Eq: BHZ-lin} results in qualitatively the same bulk spectrum, $\epsilon_{\rm bulk}^{\pm}=\pm\sqrt{ M^2+A^2k^2}$, as the BHZ model. We use units with $\hbar=e=k_B=1$ throughout the paper.

In order to describe the appearance of the edge states in the presence of boundary situated at $x=0$, we adopt the approach by Volkov and Pankratov \cite{VolkovPankratov}. We assume that inside the 2D TI ($x<0$) the mass $M$ in the Hamiltonian \eqref{Eq: BHZ-lin} is negative, whereas outside ($x>0$) the mass is equal to $+\infty$. Then, as shown in Ref.~ \cite{VolkovPankratov}, 
a pair of edge states appears as the solution of the Schr\"odinger equation with the Hamiltonian $H(-i\partial_x, k_y)$. These edge states are connected by time-reversal symmetry, and for a given $k_y$ have the following form: 
\begin{equation}
\begin{split}
\psi_{\mathrm{edge},\uparrow}(k_y, \bm{r}) & =(
1\ i\ 0\ 0)^T\frac{e^{ik_y y}}{\sqrt{2\pi \xi}}e^{-|x|/\xi} \theta(-x), \\
\psi_{\mathrm{edge},\downarrow}(k_y, \bm{r}) & =
(0\ 0\ 1\ -i)^T\frac{e^{ik_y y}}{\sqrt{2\pi \xi}}e^{-|x|/\xi}  \theta(-x),
\end{split}
\label{eq:edge:states}
\end{equation}
where  $\xi = A/|M|$ denotes the characteristic width of the edge states, and $\theta(x)$ stands for the Heaviside step function.  
The effective $2 \times 2$ Hamiltonian for the edge states can be obtained by projection of the $4\times 4$ Hamiltonian\eqref{Eq: BHZ-lin} {onto the edge states subspace} \eqref{eq:edge:states}:
\begin{gather}
H_\mathrm{edge}^{\rm (0)} = -A k_y\sigma_z .
\label{eq:edge:ham:0}
\end{gather}
The Pauli matrices $\sigma_x,\sigma_y$, and $\sigma_z$ operates in the basis of edge states.
The Hamiltonian \eqref{eq:edge:ham:0} gives rise to linear dispersion, $\epsilon^\mathrm{(edge)}_{\uparrow / \downarrow}(k_y)=\mp A k_y$.

After the projection onto the subspace of 2D electron and hole states $\vert E_1,+\rangle$, $\vert H_1,+\rangle$, $\vert E_1,-\rangle$, $\vert H_1,-\rangle$, the Hamiltonian of a magnetic impurity with spin $S$ situated at some point $\{x_0,y_0,z_0\}$ within the (001) quantum well can be written as follows  \cite{Kimme2016,Kurilovichi2016}: 
\begin{equation}
\mathcal{H}_{\rm imp} = \mathcal{J}_{\rm bulk}  \delta(x-x_0) \delta(y-y_0) ,
\end{equation}
where the matrix $\mathcal{J}_{\rm bulk}$ reads ($S_\pm = S_x\pm i S_y$):
\begin{equation}
\mathcal{J}_{\rm bulk} =
\begin{pmatrix}
J_1 S_z & -iJ_0 S_+ & J_{m} S_{-} &0\\
iJ_0S_{-} & J_2 S_z &0 &0\\
J_{m} S_{+} &0 & -J_1 S_z & - iJ_0 S_{-}\\
0 & 0 &  iJ_0 S_+ & -J_2 S_z   
\end{pmatrix} .
\label{eq:Jmat}
\end{equation}
The coupling constants $J_0$, $J_1$, $J_2$ and $J_m$ depend on $z_0$ through the envelope functions of spatially quantized states in the QW (see Ref. \cite{Kurilovichi2016} for the details). $J_0$, $J_1$, $J_2$ and $J_m$ are not necessarily positive in general. It is worthwhile to emphasize that the locations of the $S_\pm$ operators in the matrix $\mathcal{J}_{\rm bulk}$ are related to the structure of 
$\vert E_1,\pm\rangle$ and $\vert H_1,\pm\rangle$ states: The states $\vert E_1,\pm \rangle$ correspond to the $z$ component of the total angular momentum being equal to $\pm 1/2$, respectively, whereas for the states $\vert H_1,\pm\rangle$ 
this component is equal to $\pm 3/2$.

The effective interaction between the edge states and a nearby magnetic impurity can be described by means of an effective $2\times 2$ Hamiltonian, the result of projecting Eq. \eqref{eq:Jmat} onto the edge states subspace \eqref{eq:edge:states}: 
\begin{equation}
H_\mathrm{edge}^{\mathrm{(imp)}}  = \frac{1}{2\nu} \mathcal{J}_{ij}S_i \sigma_j \delta(y-y_0)
,
\label{eq: JijH}
\end{equation}
where the dimensionless coupling constants are given by the following matrix
\begin{equation} 
 \mathcal{J}  =
\frac{2\nu}{\xi}e^{-2|x_0|/\xi}\begin{pmatrix}
J_m & 0 & 2J_0\\
0 & J_m & 0\\
0 & 0 & J_z  
\end{pmatrix} . 
\label{eq: Jij}
\end{equation}
Here $\nu={1}/{(2\pi A)}$ denotes the density of states for a single helical mode and $J_z = J_1 + J_2$. The Hamiltonian \eqref{eq: JijH} -- \eqref{eq: Jij} has been derived in Refs.\cite{Otten2013,Kimme2016}. 

The structure of the matrix $\mathcal{J}$ is closely related to the structure of the bulk exchange matrix \eqref{eq:Jmat}, as well as to the structure of bulk and edge states. For example, it forbids  nonzero values of 
$\mathcal{J}_{xy}$,  $\mathcal{J}_{zy}$,  $\mathcal{J}_{yx}$, and $\mathcal{J}_{yz}$. The most striking feature of 
$H_\mathrm{edge}^{\mathrm{(imp)}}$ is the existence of the nonzero value for the element $\mathcal{J}_{xz}$. 
It allows for processes such as $|S_z=S,\:\sigma_z/2=+1/2\rangle \rightarrow |S_z=S-1,\:\sigma_z/2=+1/2\rangle$, that {do not conserve the $z$ component of {the effective} angular momentum. These transitions are allowed because the edge states \eqref{eq:edge:states} are superpositions of $|E_1,\pm\rangle$ states with $z$ component of the angular momentum being equal to $\pm 1/2$, respectively, and $|H_1,\pm\rangle$ states for which this component equals $\pm 3/2$.}

We mention that the Hamiltonian \eqref{Eq: BHZ-lin} does not take into account a possible reduction of the rotational symmetry in the  $x$-$y$ plane. Time-reversal symmetry allows for non-zero off-diagonal terms in the bulk Hamiltonian due to the presence of bulk \cite{Dai2008,Konig2008,Winkler2012,Weithofer-Recher} or interface \cite{Tarasenko2015,Durnev2016} inversion asymmetry. These off-diagonal terms result in a splitting of the electron and heavy-hole states, which has been measured recently in CdTe/HgTe/CdTe \cite{Minkov2013,Minkov2016}. Furthermore, the presence of the off-diagonal terms 
leads to a modification of the edge states \cite{Konig2008,Durnev2016} and, consequently, the matrix $\mathcal{J}$ acquires a general form with no vanishing elements \cite{Future}. 

\textsf{One-loop renormalization.}\ --- The Hamiltonian \eqref{eq: JijH} is a typical Hamiltonian for an anisotropic Kondo problem (see Ref. \cite{Zawadowski} for a review). It was derived at the energy scale of the order of the bulk gap $|M|$. Since below we are interested in edge transport which occurs at energy scales $\max\{T,|V|\}\ll |M|$, we need to take into account the renormalization of the exchange Hamiltonian \eqref{eq: JijH}. 

The interaction between the edge electrons and the magnetic impurity modifies the structure of the interaction matrix $\mathcal{J}$, leading to the Kondo effect at low temperatures. We assume that the exchange interaction is weak, $|\mathcal{J}_{jk}| \ll 1$. Then renormalization of the exchange matrix $\mathcal{J}$ can be described within the one-loop renormalization group (RG) equations. For a general form of the matrix $\mathcal{J}$ and arbitrary spin they become \cite{Zawadowski}
\begin{equation}
\frac{d\mathcal{J}_{jk}}{dl}=\frac{1}{2}\varepsilon_{jnm} \varepsilon_{kps} \mathcal{J}_{np}\mathcal{J}_{ms} .
\label{eq: RG}
\end{equation}
Here $l=\ln (|M|/E)$ is the running RG logarithmic scale and  $\varepsilon_{jnm}$ stands for the Levi-Civita symbol. The band gap $|M|$ serves as the ultraviolet cutoff whereas $E=\max\{T,|V|\}$ determines the infrared energy scale. Using 
Eqs. \eqref{eq: RG}, one can find the following equations,
\begin{equation}
\frac{d(\mathcal{J} \mathcal{J}^T)_{jk}}{dl}=\frac{d(\mathcal{J}^T \mathcal{J})_{jk}}{dl}=2  \delta_{jk} \det \mathcal{J} .
\label{eq:inv}
\end{equation}
It is convenient to perform the singular value decomposition of the initial matrix, $\mathcal{J}(l=0)= R_{<} \Lambda R_{>}$, where the $SO(3)$ matrices $R_{<,>}$ do not flow, and where $\Lambda= {\rm diag}(\lambda_1,\lambda_2,\lambda_3)$. Then, in the course of the RG flow, the matrix $\Lambda$ preserves its diagonal form with $\lambda_j$ satisfying the following equations
\begin{equation}
\frac{d\lambda_1}{dl} = \lambda_2\lambda_3, \quad \frac{d\lambda_2}{dl} = \lambda_1\lambda_3, \quad \frac{d\lambda_3}{dl} = \lambda_1\lambda_2 .
\label{eq: RG-2}
\end{equation}
Eqs. \eqref{eq: RG-2} have the two independent invariants $\lambda_1^2-\lambda_2^2$ and $\lambda_2^2-\lambda_3^2$. 
In the general case, if none of these invariants is zero, the RG flow tends to the manifold $|\lambda_1|=|\lambda_2|=|\lambda_3|$  with $\lambda_1\lambda_2\lambda_3>0$.
All three $\lambda_j$ diverge at a finite scale $l_K$, which determines the Kondo temperature $T_K=|M| e^{- l_K}$. 
In what follows, we assume that $\max\{T,|V|\} \gg T_K$ so that $|\mathcal{J}_{jk}|\ll 1$ at the corresponding scale.

\textsf{Backscattering current and master equation.}\ --- In order to obtain the correction to the current flowing along the 
edge  due to scattering off a magnetic impurity, we shall follow an approach developed recently in Ref. \cite{Goldstein2014}. The helicity of the edge states allows one to relate the correction to the current, $\Delta I$, to the rate of change of the $z$ component of the total spin of the edge electrons:
\begin{gather}\label{eq: c-s}
    \Delta I=\left \langle \frac{d}{dt}  \int dy\, s_z \right \rangle,\,
    s_z(y)=\frac{1}{2} \Psi^\dagger (y) \sigma_z \Psi(y) ,
\end{gather}
where $\Psi^\dagger$ and $\Psi$ denote creation and annihilation operators of  the edge electrons, respectively.

The $z$ component of the total spin of the edge electrons is not conserved due to the exchange interaction with the magnetic impurity, Eq. \eqref{eq: JijH}. To second order in the exchange interaction $\mathcal{J}_{jk}$ we find \cite{Future},
\begin{equation}
\frac{\Delta I}{G_0 V} = \frac{\pi^2}{2} \Bigl [ \mathcal{X}_r\langle S_r\rangle 
 \coth \frac{V}{2T} - 2\!\sum\limits_{k=x,y} \!\mathcal{J}_{mk} \mathcal{J}_{pk} \langle S_m S_p \rangle\Bigr ] .
\label{eq:curr:corr}
\end{equation} 
Here  
$\mathcal{X}_j = 2 \varepsilon_{jkl} \mathcal{J}_{kx}\mathcal{J}_{ly}$ and $\langle\cdots\rangle$ denotes the average with respect to the steady-state reduced density matrix $\rho_S$ of the impurity spin, e.g., $\langle S_r \rangle = \Tr S_r \rho_S$. In order to determine the steady state we derived the following equation for the density matrix of the impurity spin within second order perturbation theory in $\mathcal{J}_{jk}$ \cite{Future}:
\begin{equation}
\frac{d\rho_S}{dt} = \frac{\langle s_z \rangle \mathcal{J}_{jz}}{i\nu} \bigl [S_j, \rho_S \bigr ] +
\eta_{jk}\Bigl ( S_j  \rho_S S_k -\frac{1}{2}\{\rho_S, S_k S_j \} \Bigr ) .
\label{eq:ME:S1}
\end{equation}
Here 
the average $z$ component of the edge spin density  $\langle s_z \rangle=\nu V/2$ is evaluated disregarding the influence of the 
impurity on the distribution of the edge electrons. The matrix $\eta$ is defined as $\eta = \pi T \mathcal{J} \Pi_V \mathcal{J}^T$, where
\begin{equation}
\Pi_V  = 
\begin{pmatrix}
\frac{V}{2T} \coth \frac{V}{2T} & - i \frac{V}{2T} & 0\\
i \frac{V}{2T} & \frac{V}{2T} \coth \frac{V}{2T} & 0 \\
0 & 0 & 1
\end{pmatrix} .
\end{equation}
We note that the eigenvalues of $\Pi_V$ are equal to $1$ and $(V/2T) [\coth (V/2T) \pm 1] \geqslant 0$. Therefore, 
the matrix $\eta$ is positive semidefinite and the master equation \eqref{eq:ME:S1} has the Lindblad form, ensuring the positivity of $\rho_S$ \cite{BreuerPetruccione}. 
The vector $\langle s_z \rangle \mathcal{J}_{jz}$ plays the role of the effective magnetic field in which the impurity spin rotates, whereas $\eta_{jk}$ is responsible for Korringa-type relaxation. 

The expression for the correction to the current \eqref{eq:curr:corr} and the master equation \eqref{eq:ME:S1} are invariant
under rotation of the exchange matrix $\mathcal{J}$ from the left by an arbitrary orthogonal matrix $U$, $\mathcal{J} \to U \mathcal{J}$. Indeed, since the following relation holds, $\det \mathcal{J} = \mathcal{X}_j\mathcal{J}_{jz}/2$, the vector $\mathcal{X}$ transforms as follows: $\mathcal{X} \to U \mathcal{X}$. 

Using the steady state solution of the Lindblad equation \eqref{eq:ME:S1} one can compute the averages in Eq. \eqref{eq:curr:corr} and find 
$\Delta I$. For example, multiplying both sides of Eq. \eqref{eq:ME:S1} by $S_q$ and taking the trace, we derive the following equation for the average impurity spin:
\begin{gather}
\frac{d\langle S_q \rangle}{dt} = \frac{\pi T}{2} \Biggl [ \frac{V}{2T} 
\varepsilon_{zkl} \varepsilon_{qmr}  \mathcal{J}_{mk} \mathcal{J}_{pl}
\langle \{S_r, S_p\} \rangle - \Gamma_{qr} \langle S_r\rangle 
 \Biggr ]  ,
 \label{eq:ME:spin}
\end{gather}
where we introduced the following $3\times 3$ matrix 
\begin{gather}
\Gamma_{qr} = \frac{1}{\pi T} \Bigl[ \delta_{qr} \Tr \eta  - \frac{\eta_{qr}+\eta_{rq}}{2}
+V \varepsilon_{qrj}\mathcal{J}_{jz}
\Bigr ] .
\end{gather}
In general it is not an easy task to find the averages analytically, since $\langle S_r\rangle$ is related to $\langle \{S_m, S_p\} \rangle$, etc.

\textsf{Linear conductance.}\ --- In spite of the complicated structure of the master equation \eqref{eq:ME:S1} for an arbitrary value of the impurity spin $S$, the correction to the linear conductance, $\Delta G = \lim\limits_{V\to 0} {\Delta I}/{V}$, 
can be found analytically.  As one can see from Eq. \eqref{eq:curr:corr}, it is enough  to compute $\langle \{S_r, S_p\} \rangle$ for $V=0$, and to find $\langle S_r \rangle$ to the first order in $V$. Using {the fact that} at zero bias voltage $\langle \{S_r, S_p\} \rangle = 2S(S+1)\delta_{rp}/3$, we find from Eq. \eqref{eq:ME:spin}:
\begin{equation}
\langle S_{r}\rangle = \frac{S(S+1)}{3}\frac{V}{T} (\Gamma_0^{-1})_{rq} \mathcal{X}_q .
\end{equation}
Here we introduce $\Gamma_0 = \Tr (\mathcal{J}\mathcal{J}^T) - \mathcal{J}\mathcal{J}^T$, which is the matrix $\Gamma$ evaluated at $V=0$. Then, from Eq. \eqref{eq:curr:corr} we obtain the following correction to the conductance
\begin{gather}
\frac{\Delta G}{G_0}  =  \frac{\pi^2 S(S+1)}{3} \Bigl [ \mathcal{X}^T \Gamma_0^{-1}\mathcal{X}  - 
g
\Bigr ] ,
\label{eq:curr:s0}
\end{gather}
where $g = (\mathcal{J}^T \mathcal{J})_{xx}+ (\mathcal{J}^T \mathcal{J})_{yy}$. Interestingly, the impurity spin 
enters 
$\Delta G$
as an overall factor $S(S+1)$ only. 

The result \eqref{eq:curr:s0} can be expressed via the matrix $R_>$ and the parameters $\lambda_j$ as follows:
\begin{equation}
\label{eq: RGconductance-2}
\frac{\Delta G}{G_0} = - \frac{\pi^2 S(S+1)}{3} \left [ R_{>}^{-1} \Phi R_{>}
\right ]_{zz} ,
\end{equation}
where 
\begin{equation}
\Phi = {\rm diag} \left (
\frac{(\lambda_2^2-\lambda_3^2)^2}{\lambda_2^2+\lambda_3^2}, \frac{(\lambda_1^2-\lambda_3^2)^2}{\lambda_1^2+\lambda_3^2}, \frac{(\lambda_1^2-\lambda_2^2)^2}{\lambda_1^2+\lambda_2^2} \right ) .
\end{equation}
{As one can see, in the case of the exchange matrix of the form $\mathcal{J}^{\rm(iso)}_{jk} = \{\mathcal{J}_\perp, \mathcal{J}_\perp, \mathcal{J}_z\}$, $\Delta G$ is exactly zero \cite{Tanaka2011}. Therefore, the correction is zero  for any exchange matrix of the form $U \mathcal{J}^{\rm(iso)}$. 
If the matrix does not have this form, the correction to the conductance \eqref{eq: RGconductance-2} is negative, $\Delta G < 0$. Hence, the scattering off a magnetic impurity results in suppression of the conductance of the helical edge. 
The correction to the conductance \eqref{eq: RGconductance-2} is independent of temperature except for a weak temperature dependence due to the renormalization of $\lambda_j$ [Eq. \eqref{eq: RG-2}]. Since the matrix $R_>$ {is fixed} and the differences $\lambda_j^2-\lambda_k^2$ are RG invariants while $|\lambda_j|$ eventually grow, $\Delta G$ drops to zero as 
$T$ approaches $T_K$.
We note that for spin $S=1/2$ and for the exchange matrix that slightly deviates from $\mathcal{J}^{\rm(iso)}$ our result \eqref{eq:curr:s12} reproduces the results of Ref. \cite{Goldstein2014}.}
If one neglects the renormalization of $\mathcal{J}_{jk}$, one may plug $\mathcal{J}$ from Eq. \eqref{eq: Jij} into Eq. \eqref{eq: RGconductance-2} to find 
\begin{equation}
\label{eq: RGconductance-f}
\frac{\Delta G}{G_0} = - \frac{4 S(S+1)}{3}\frac{M^2}{A^4} \frac{J_0^2J_m^2 e^{-4|x_0|/\xi}}{J_m^2+J_z^2+2J_0^2} .
\end{equation}

\textsf{Backscattering current for spin $S=1/2$.}\ --- In the case of a spin $S=1/2$ impurity we can find the backscattering current at arbitrary values of $V$. Since $\{S_r, S_p\} = \delta_{rp}/2$ for spin $S=1/2$, Eq. \eqref{eq:ME:spin} leads to the stationary solution $\langle \bm{S}_r\rangle = V (\Gamma^{-1})_{rq} \mathcal{X}_q/(4T)$. Simultaneously, the correction to the current, Eq. \eqref{eq:curr:corr}, is expressed via $\langle \bm{S}\rangle$ only. Finally, we find
\begin{gather}
\Delta I =  G_0 \frac{\pi^2 V}{4} \Bigl [ \mathcal{X}^T \Gamma^{-1}\mathcal{X} \frac{V}{2T} \coth \frac{V}{2T} \,    - g
 \Bigr ] .
\label{eq:curr:s12}
\end{gather}
For an exchange matrix of the form $U \mathcal{J}^{\rm(iso)}$ the backscattering correction to the current is zero at any voltage \cite{Tanaka2011}.

\textsf{Backscattering current for an arbitrary spin.}\ --- In the case of spin $S>1/2$ the master equation \eqref{eq:ME:S1} cannot be reduced to a closed equation for the average spin. However, at large voltage, $|V| \gg \max\{|\mathcal{J}_{jk}|\} T$, one can find the stationary solution of the master equation \eqref{eq:ME:S1} in the following way. 
In this regime the first term in the right hand side of Eq. \eqref{eq:ME:S1} dominates over the second one. Let us introduce the effective Hamiltonian $\mathcal{H}_V = - \langle s_z\rangle \mathcal{J}_{jz} S_j/\nu$. Then, it is 
reasonable to look for a stationary density matrix which commutes with $\mathcal{H}_V$, i.e., which is diagonal in eigenbasis of $\mathcal{H}_V$. The right hand side of the Lindblad equation \eqref{eq:ME:S1} then yields
\begin{equation}
\rho_{S,l} = \sum_m \frac{\eta_{jk} \langle l | S_j |m\rangle \rho_{S,m} \langle m | S_k |l \rangle}{\eta_{jk} \langle l | S_k S_j |l \rangle} .
\label{eq:HV:sol}
\end{equation}
{Here $|l\rangle$ and $|m\rangle$ denotes eigenstates of $\mathcal{H}_V$, i.e. states with a given angular momentum projection in the direction of the effective magnetic field mediated by the interaction with the electrons, while $\rho_{S,m} =\langle m |\rho_S| m\rangle$.} Since 
the Hamiltonian $\mathcal{H}_V$ is linear in the spin operators, its eigenenergies are linear functions of $l$. Furthermore, the intermediate states $m$ in Eq. \eqref{eq:HV:sol} are equal to $l$ or $l \pm 1$ only. Hence, it is possible to solve Eq. \eqref{eq:HV:sol} by the Gibbs ansatz \cite{Future}:
$\rho_S = \exp(-\mathcal{H}_V/T_{\rm eff}) /\Tr \exp(- \mathcal{H}_V/T_{\rm eff})$.
Then, upon summation over $l$, Eq. \eqref{eq:HV:sol} becomes equivalent to the following equation:
\begin{equation}
\eta_{jk} \Tr \bigl (  S_k S_j \bigr )= \eta_{jk} \Tr \Bigl ( e^{ \mathcal{H}_V/T_{\rm eff} } S_j  e^{- \mathcal{H}_V/T_{\rm eff} } S_k \Bigr ) .
\label{eq:RRR-1}
\end{equation}
In order to find the effective temperature $T_{\rm eff}$, it is convenient to define the matrix $\mathcal{C}_{jk} = i \varepsilon_{jkl} \langle s_z \rangle \mathcal{J}_{lz}/\nu$. Then Eq. \eqref{eq:RRR-1} can be equivalently rewritten as 
$\Tr \eta = \Tr \bigl [\mathcal{R} ^{-1} \eta \mathcal{R}\, \exp \bigl (\mathcal{E}_V/T_{\rm eff} \bigr )\bigr ]$, 
where $\mathcal{E}_V = {\rm diag} (1,0,-1) V\sqrt{(\mathcal{J}^T\mathcal{J})_{zz}}/2$ and $\mathcal{C} =  \mathcal{R} \mathcal{E}_V \mathcal{R} ^{-1}$. The 
effective temperature reads
\begin{equation}
 T_{\rm eff} =  \frac{V}{4} \sqrt{(\mathcal{J}^T\mathcal{J})_{zz}} \Biggl \{
  {\rm arccoth \,} \frac{(\mathcal{J}^T\Gamma_0\mathcal{J})_{zz}\coth\frac{V}{2T}}{2(\det \mathcal{J}) \sqrt{(\mathcal{J}^T\mathcal{J})_{zz}}}
\Biggr \}^{-1} .
\label{Eq:Teff:1}
\end{equation}
The effective temperature $T_{\rm eff}$ therefore depends on the voltage $V$, the temperature $T$, and the exchange matrix $\mathcal{J}$, and is of order  $|\mathcal{J}_{jk}|\max\{T,|V|\}$. Interestingly, the effective temperature is independent of the value of the spin $S$.
Moreover, the Gibbs factor $\exp( -\mathcal{H}_V/T_{\rm eff})$ is independent of the voltage at $|V|\gg T$. Therefore, at $|V|\gg T$ the impurity spin is not fully polarized. 
We note that the result \eqref{Eq:Teff:1} can be easily expressed via the matrix $R_>$ and parameters $\lambda_j$.

\begin{figure}[t]
\centerline{\includegraphics[width=0.52\textwidth]{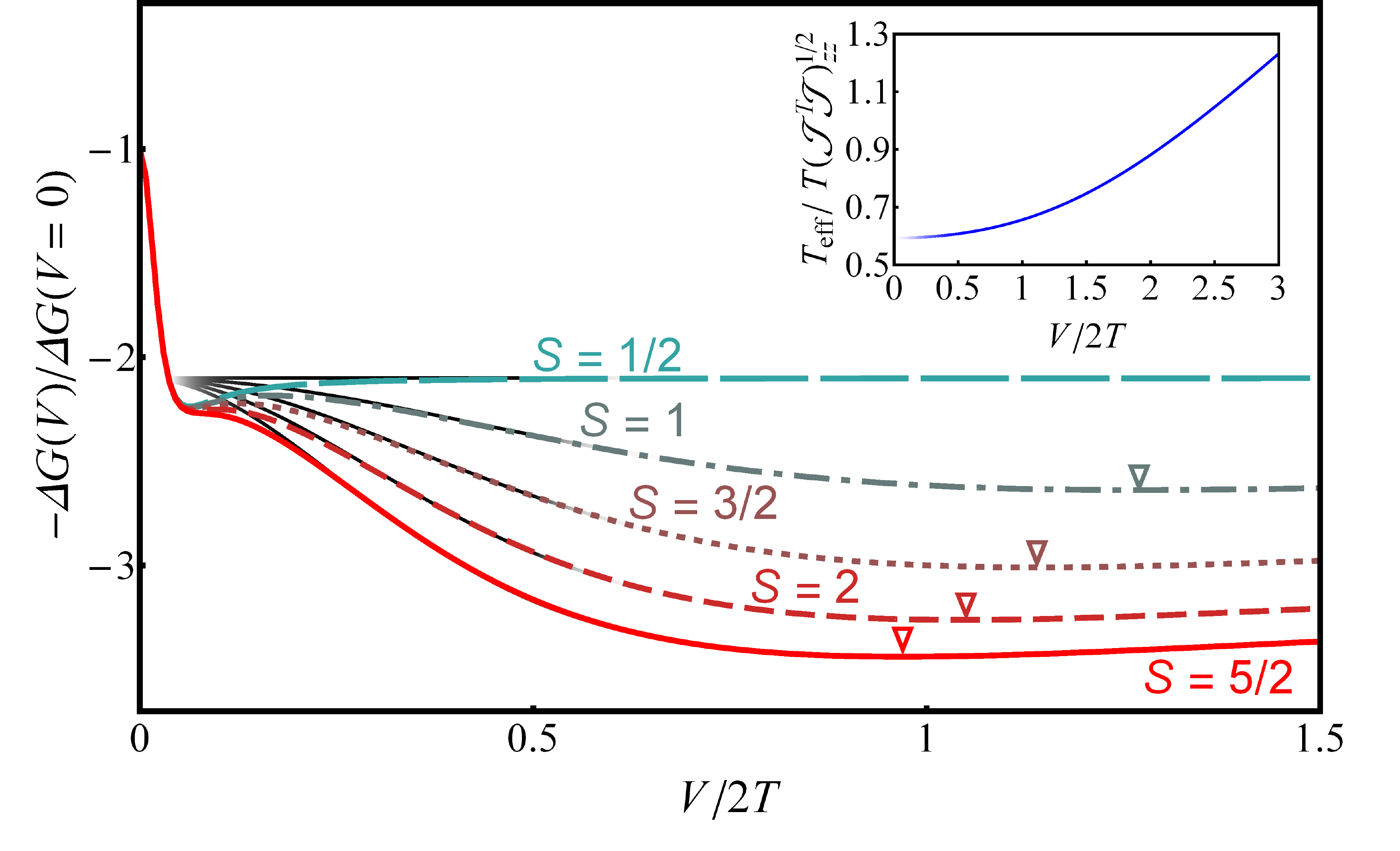}}
\caption{Fig. \protect\ref{Figure2}. $-\Delta G(V)/\Delta G(0)$ versus $V/2T$ for different values of $S$. The exchange couplings were chosen as $\mathcal{J}_{xx}=\mathcal{J}_{yy}=10^{-2}$, $\mathcal{J}_{xz}=0.8 \mathcal{J}_{xx}$, $\mathcal{J}_{zx}=0.3\mathcal{J}_{xx}$, $\mathcal{J}_{zz}=0.9\mathcal{J}_{xx}$, while the other couplings vanish. Black thin curves correspond to the approximate solutions with $\rho_S$ determined by $T_{\rm eff}$ (see Eq. \eqref{Eq:Teff:1}). The empty triangles indicate the positions of additional minima.  \color{black}
Inset: the dependence of $T_{\rm eff}$ on $V$. 
}
\label{Figure2}
\end{figure}

If one neglects the renormalization of $\mathcal{J}_{jk}$, then for  $\mathcal{J}_{jk}$ given by Eq. \eqref{eq: Jij} the effective temperature acquires the following simple form ($J=\sqrt{4{J}_{0}^2+{J}_{z}^2}$):
 \begin{equation}
 T_{\rm eff} =   \frac{J V |M|e^{-2|x_0|/\xi}}{4\pi A^2} \Bigl \{
{\rm arccoth \,} \Bigl [\frac{J^2+{J}_{z}^2}{2 J {J}_{z}} \coth \frac{V}{2T}
\Bigr ]  \Bigr \}^{-1}.
\label{Eq:Teff:1-3}
 \end{equation}
We note that our result \eqref{Eq:Teff:1-3} is different from the result of Ref. \cite{Kimme2016} in which the stationary density matrix was determined in the case of the exchange matrix \eqref{eq: Jij} at $|V|\gg \max\{|\mathcal{J}_{jk}|\} T$. The solution obtained in Ref. \cite{Kimme2016} leads to the full polarization of the impurity spin at $|V|\gg T$. 
This conclusion is a consequence of the assumption made in Ref. \cite{Kimme2016} that $\rho_S$  is diagonal in the eigenbasis of $S_z$ rather than $\mathcal{H}_V$, an assumption which is not justified in general.

Having found the effective temperature, one can use the stationary Gibbs-like density matrix
to
compute the averages $\langle S_r\rangle $ and $\langle S_m S_p \rangle$ appearing in Eq. \eqref{eq:curr:corr}. In this way, one can determine the correction to the current in the regime of large voltage, $|V|\gg \max\{|\mathcal{J}|_{jk}\} T$. 
For spin $S=1/2$ we reproduce the result \eqref{eq:curr:s12} in this regime.

An example of the \color{black} dependence of the backscattering correction to the differential conductance $\Delta G(V)$ on the voltage for the different values of the impurity spin $S$  is shown 
in Fig.~\ref{Figure2}. The curves are obtained by numerical solution of the master equation \eqref{eq:ME:S1} without taking into account the renormalization of $J_{jk}$. 
For $S>1/2$ the differential conductance is non-monotonous, with extrema at $V\sim |\mathcal{J}_{jk}| T$ and $V\sim T$  (indicated by triangles in Fig. ~\ref{Figure2}). The
first extremum is the consequence of~com\-pe\-ti\-tion between the effective magnetic field acting on the impurity spin 
 and  the relaxation (the first and second terms in the right hand side of Eq. \eqref{eq:ME:S1}, respectively). The extremum at $V\sim T$ is the consequence of the dependence of the effective temperature on the voltage.
\color{black}
In contrast with the higher spins, for $S=1/2$ the differential conductance saturates  already \color{black} at $V \sim |\mathcal{J}_{jk}| T$ instead of $V \sim T$. This feature follows directly from 
the analytical solution
\eqref{eq:curr:corr}. 
Finally, we mention that in the case of $|V| \ll |\mathcal{J}_{jk}|T$ the ratio $\Delta G (V)/\Delta G(0)$ is independent of $S$.

\textsf{Conclusions.}\ --- In conclusion, we presented the results of a detailed study of the backscattering current at a helical edge due to weak scattering off a single Kondo-type magnetic impurity. Contrary to the previous studies 
we considered the case of a magnetic impurity with an arbitrary spin $S$ and a general exchange matrix. For $S=1/2$ we found an analytical expression \eqref{eq:curr:s0} for the backscattering current valid at arbitrary voltage. For spin $S>1/2$ we found analytical expressions for the backscattering current at low and high voltages.

For a (001) CdTe/HgTe/CdTe QW with width close to the critical one, $6.3$ nm, the exchange couplings
$|J_m|, |J_z|, |J_0|$ for a Mn impurity can be estimated to be of the order of $0.1$ $\mathrm{eV\cdot nm^2}$  \cite{Kurilovichi2016}. 
Using the estimates $\nu \simeq 0.5\: \mathrm{eV^{-1}\cdot nm^{-1}}$ and  $\xi \simeq 40$ nm \cite{Qi-Zhang}, we find that the exchange couplings $\mathcal{J}_{jk}$ are of the order $10^{-3}$. This implies that the backscattering correction to the conductance due to a single Mn impurity is of order $-{\Delta G}/{G_0}\sim 10^{-4} \div 10^{-3}$.

Finally, we mention that in the case $S>1/2$ a local anisotropy term for the impurity spin is generated \cite{RKonig,Schiller}, e.g., due to indirect exchange interaction mediated by the bulk and edge states. The corresponding anisotropy Hamiltonian strongly  affects the dynamics of the impurity spin at low temperature and voltage and, consequently, changes the results for the edge transport \cite{Future}.  

We thank Yuval Gefen  for fruitful collaboration in the initial stage of this project and for very useful discussions.
I.S.B. is grateful to Weizmann Institute of Science and Tel Aviv University for their hospitality. The work was partially supported by the Russian Foundation for Basic Research under the Grant No. 15-52-06005, Russian President Grant No. MD-5620.2016.2, Russian President Scientific Schools Grant NSh-10129.2016.2, the Ministry of Education and Science of the Russian Federation under the Grant No. 14.Y26.31.0007, the Alexander von Humboldt Foundation,
the Israel Ministry of Science and Technology (Contract 3-12419), the Israel Science Foundation (Grant 227/15), the German-Israeli Foundation (Grant I-1259-303.10), and the US-Israel Binational Science Foundation (Grant 2014262).

\end{document}